\documentclass[a4paper,11pt]{article}
\usepackage{graphicx}
\usepackage{bbm}
\usepackage[english]{babel}
\usepackage{amsmath}
\newcommand\e{{\rm e}}
\renewcommand\i{{\rm i}}

\begin{document}

\title{Quantum search algorithms on the hypercube} 
\author{Birgit Hein and Gregor Tanner\\
\it School of Mathematical Sciences, University of Nottingham, \\
\it University Park, Nottingham NG7 2RD, UK\\
\it {\rm e-mail:} gregor.tanner@nottingham.ac.uk}
\date{17th June 2009}
\maketitle
{J.\ Phys.\ A:\ Math.\ Theor.\ {\bf 42} (2009) 085303 }

\begin{abstract}
We investigate a set of discrete-time quantum search algorithms on 
the n-dimensional hypercube following a proposal by Shenvi, 
Kempe and Whaley \cite{SKW03}. We show that there exists 
a whole class of quantum search algorithms in the symmetry reduced
space which perform a search of a marked vertex in time of order
$\sqrt{N}$ where $N = 2^n$, the number of vertices.
In analogy to Grover's algorithm, the spatial search is  
effectively facilitated through a rotation in a two-level sub-space 
of the full Hilbert space. In the hypercube, these two-level
systems are introduced through avoided crossings. We give estimates on 
the quantum states forming the 2-level sub-spaces at the 
avoided crossings and derive improved estimates on the search times. 
\end{abstract}

\section{Introduction}
The recent interest in quantum random walks and quantum search 
algorithms is fuelled by the discovery that wave systems can perform
certain tasks more efficiently than classical algorithms. Leaving aside
issues about the implementation of such algorithms, it is essentially 
wave interference which acts here as an additional resource. The prime 
example is {\em Grover}'s search algorithm \cite{Gro96, NC00} which 
performs a search in a data-base of $N$ items in $\sqrt{N}$ steps whereas
classical search algorithms need of order $N$ steps. 

In Grover's algorithm it is assumed that one can act on all data points
directly. In many physical situations, interactions between 
points in the network may be restricted and the search can only take place 
between nodes which are directly connected. Classical search algorithms 
are then typically based on (weighted) random walks exploring the network 
through jumps from one node to another with prescribed probabilities. The quantum 
analogue  - a quantum random walk - has been introduced recently. It can be 
shown that for certain network topologies improvements on
transport speed or hitting times compared to classical random walks 
can be achieved - see \cite{kempe, Amb03} for overviews. The connection 
between quantum random walks and quantum graph theory \cite{SS06} 
has been pointed out in \cite{ST04, Tan07}.

In \cite{AA03, AKR05} quantum random walk algorithms have been introduced 
which lead to speed-up in a spatial search on an $n$-dimensional regular grid; 
a continuous-time version of such a quantum search algorithm has been given in 
\cite{CG04}. In \cite{SKW03}, an algorithm for a search on the $n$ dimensional
hypercube has been presented; we will focus on this algorithm in what
follows, reinterpreting and generalising the results in \cite{SKW03} as well
as giving improved estimates for the search time.\\

The discrete time search algorithms listed above have 
in common that they are based on quantum walks on regular 
(quantum) graphs \cite{ST04}. 
The quantum walk itself is defined by a regular graph, a 'quantum coin flip'
(equivalent to a local scattering matrix at each vertex), and a 
'shift operation' \cite{kempe, Tan07}. A marked vertex is then introduced 
in the form of a local perturbation of the quantum random walk.
Typically, this is achieved by modifying the coin flip at this vertex; the 
quantum search algorithm will localise at the target vertex after 
$T$ time steps.

This paper is organised as follows: first we will introduce the algorithm and 
give the eigenvalues of the unperturbed walk in terms of a unitary propagator
$U$. Then we will introduce a symmetry reduced space and proceed to the 
spectrum of a quantum search algorithm $U_{\lambda}$; here, $\lambda$ 
characterises the perturbation strength at the marked vertex extrapolating 
between the unperturbed walk ($\lambda = 0$) and a maximal perturbation 
($\lambda = 1$); the latter corresponds to the search algorithm 
in \cite{SKW03}. The spectrum of 
$U_{\lambda}$ shows several avoided crossings - each of these
crossings can be used to construct a search algorithm in the reduced 
space. We will give estimates for the vectors spanning the two-level subsystems
at each crossing and derive the leading asymptotics for the search times. 
We show in particular that the search algorithm for the central crossing 
at $\lambda = 1$ finds the marked vertex in 
$T\approx\pi \sqrt{N\left(\frac{1}{8}+\frac{1}{32n}\right)}$ time 
steps where $N = 2^n$ and $n$ is the dimension of the hypercube. We 
will return to the search on the full hypercube space in the last 
section.

\section{The Shenvi-Kempe-Whaley search algorithm on the hypercube}

\begin{figure}
\centering
\includegraphics[scale=0.3]{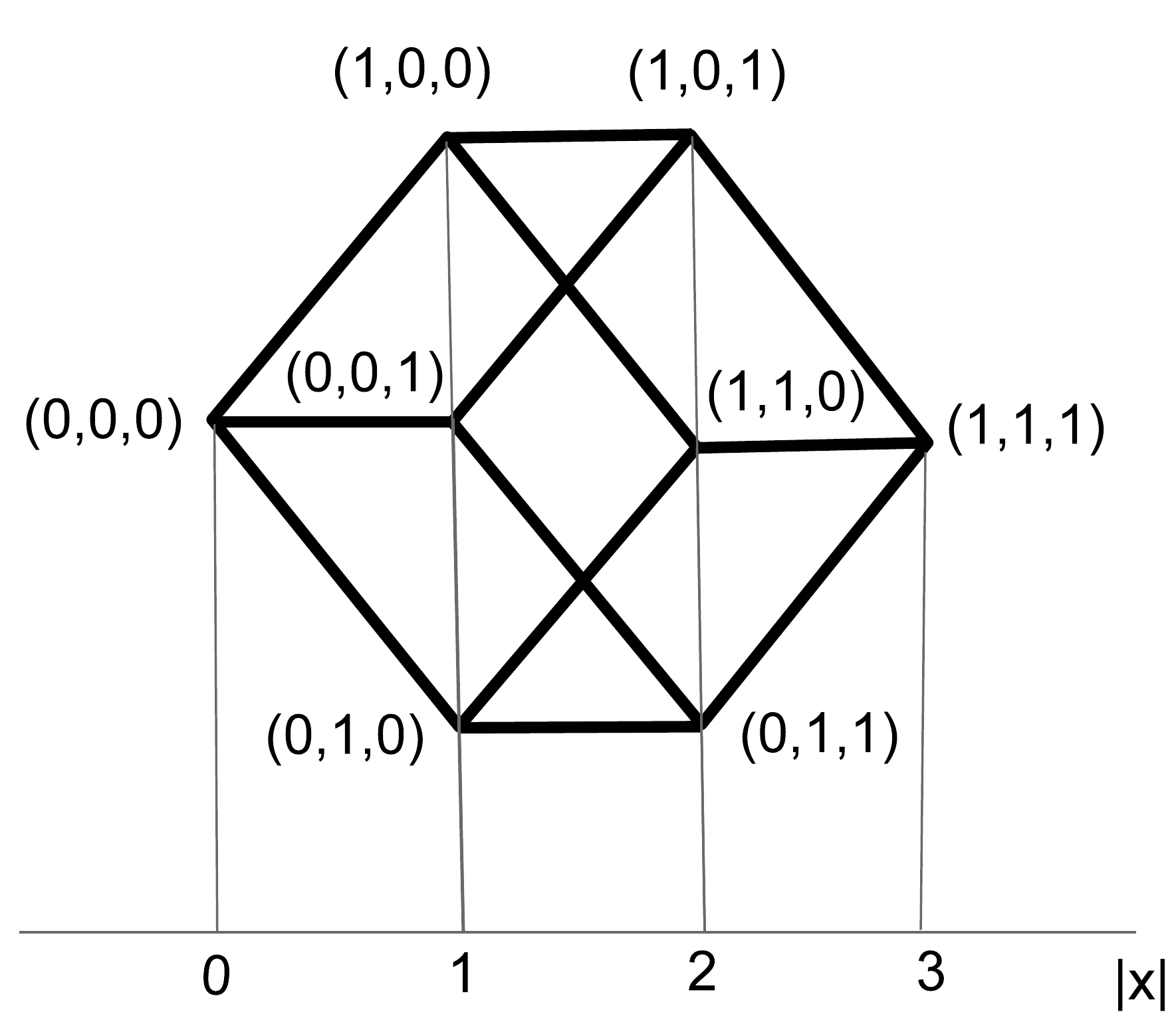}
\caption{The hypercube in $n=3$ dimensions. The Hamming weight 
$\left|\vec{x}\right|$ measures the distance 
between $\vec{x}$ and $\vec{0}$. To illustrate this, all vertices with 
the same Hamming weight are projected to one point.}
\label{fig:3-d hypercube}
\end{figure}

The vertices of the $n$-dimensional hypercube can be encoded using binary 
strings with $n$ digits. All vertices whose strings are equal for all but 
one digit are connected, as shown in Fig.~\ref{fig:3-d hypercube}. Hence, 
each vertex is connected to $n$ neighbouring vertices. Since the position 
space has $2^{n}$ dimensions and the coin space has $n$ dimensions, the 
Hilbert space $\cal H$ is of size $2^{n}n$. We denote the 
corresponding unit vectors in the position representation as 
$\left|d,\vec{x}\right\rangle$ with $d = 1,\ldots,n$ specifying the 
direction at each vertex and $\vec{x}$ 
gives the coordinates of the vertex, see Fig.\ \ref{fig:3-d hypercube}.

We will start from the quantum walk previously defined in \cite{SKW03}. 
A coin flip $C$ acting simultaneously on the internal degrees of freedom at 
all vertices defines the direction in which the walk will be shifted.  
The shift operator $S$ shifts the walk to neighbouring vertices. The 
(unperturbed) walk is then given as $U=SC$. 

The local coin flip is specified by a unitary coin matrix $C_0$ which connects  
$n$ incoming with $n$ outgoing channels at each vertex and acts effectively 
as a vertex scattering matrix. We use the uniform distribution in coin space 
$\left|s\right\rangle=\frac{1}{\sqrt{n}}\sum_{i=1}^{n}\left|i\right\rangle $ 
to define the local coin flip on each vertex as 
$ C_{0}=2\left|s\right\rangle\left\langle s\right|-\mathbbm{1}_{n}$. Using 
the tensor product and the identity matrix in position (vertex) space,  
$\mathbbm{1}_{2^{n}}$, the global coin flip is defined as 
$C=C_{0}\otimes \mathbbm{1}_{2^{n}}$.

The shift operator moves the quantum walk to one of the neighbouring vertices. 
The state in $\left|d,\vec{x}\right\rangle$ is shifted to 
$\left|d,\vec{x}\oplus\vec{e_{d}}\right\rangle$, where $\vec{e_{d}}$ is 
the unit vector in direction $d$ and hence
\begin{equation}
S=\sum_{d=1}^{n}\sum_{\vec{x}} \left| d,\vec{x}\oplus\vec{e_{d}}\right\rangle\left\langle d,\vec{x}\right|
.\end{equation}
The relevant eigenvectors and eigenvalues of $U=SC$ are
\begin{eqnarray} \label{eigenw}
v_{k}^{\pm}
&=& \e^{\pm\i \omega_{k}}
= 1-\frac{2k}{n}\pm\frac{2\i}{n}\sqrt{k\left(n-k\right)}  \\ \label{eigenv}
\vert v_{\vec k}^{\pm}\rangle 
&=&\beta_k \sum_{\vec x} \left(-1\right)^{\vec k \cdot\vec x} 
\frac{2^{-n/2}}{\sqrt{2}}\sum_{d=1}^n \alpha_{k_{d}}^{\pm} \left|  
d,\vec x\right\rangle 
\end{eqnarray}
\cite{SKW03,ref11}, where 
the vector $\vec k$ consist of $n$ entries that can take the values $0$ or $1$ 
and $k=|\vec{k}|$ denotes the Hamming weight of this vector, i.e.\ the sum 
of all entries. Furthermore,
\begin{eqnarray} \label{alpha}
\alpha_{k_{d}}^{\pm}&=&\begin{cases} 
1/\sqrt{k} & \text{if }  k_{d}=1\\ \mp \i/\sqrt{n-k}, & \text{if }  k_{d}=0
\end{cases}\\ \label{beta}
\beta_{k}&=&\begin{cases} 
\sqrt{2}, & \text{if }  k=0 \text{ or } k=n \\ 1 & \text{else } 
\end{cases}
\, ,
\end{eqnarray}
and $k_d$ is the $d$-th component of $\vec k$. Note that for $k=0$ or $n$ 
the two cases $\pm$ are equivalent. The eigenvalues $e^{i \omega_{k}}$ are 
${n\choose k}$ times degenerate. All other eigenvalues of $U$ are $\pm 1$;
the corresponding eigenvectors are not affected by the perturbation and 
are related to the spectrum of the coin space, see \cite{ref11}. We do not
need to consider this trivial eigenspace in what follows.\\
We now employ this quantum walk to construct a quantum search algorithm. We 
mark the target vertex $v$ with a different coin flip, 
that is, we choose a local coin matrix at $v$ with 
$C_{1}=-\mathbbm{1}_{n}$. The marked coin acting on the full Hilbert space
is then given as
\begin{equation}
C^{\prime}=C-\left(C_{0}-C_{1}\right)\otimes 
\left|v\right\rangle \left\langle v\right| \, .
\end{equation}
This defines a `perturbed' quantum walk $U^{\prime}=SC^{\prime}$, which can 
be written as 
\begin{equation} \label{uprime}
U^{\prime}=U\left(\mathbbm{1}_{2^{n}n}- 2
\left|sv\right\rangle \left\langle sv\right|\right)\, , 
\end{equation}
where $\left| sv\right\rangle=\left| s\right\rangle\otimes\left| v\right\rangle$
is a state localised at the marked vertex and uniformly distributed in coin 
space. Note that $|sv\rangle$ is orthogonal to all trivial eigenvectors \cite{ref11}. 
Eqn.\ (\ref{uprime}) is obtained using the definition of $C_0$ and $C_1$, 
that is,
\begin{eqnarray}
\label{Eigenvectors of U}
U^{\prime}&=&SC^{\prime}=
SC-S\left(\left(C_{0}-C_{1}\right)\otimes \left|v\right\rangle 
\left\langle v\right|\right)\nonumber\\
&=& SC-S\left(\left(2|s\rangle\langle s|-\mathbbm{1}_{n}
+ \mathbbm{1}_{n}\right)\otimes |v\rangle \langle v| \right) \nonumber\\
&=& SC-2SC\left(|s\rangle\langle s|\otimes |v\rangle \langle v|\right) 
\nonumber\\
&=&U\left(\mathbbm{1}_{2^{n}n}- 2|sv\rangle 
\langle sv|\right)
,\end{eqnarray}
where we use $C_0|s\rangle = |s\rangle$.
The search algorithm is now started in the eigenstate 
$\left|v_{\vec 0}\right\rangle$ of the unperturbed walk
which is uniformly distributed over the whole Hilbert space, see 
Eqn.\ (\ref{eigenv}). It has been shown in \cite{SKW03} that the state
$\left(U'\right)^t |v_{\vec 0}\rangle$ 
localises on the marked vertex after $t = {\cal O}(\sqrt{N})$ steps
with $N = 2^n$, the total number of vertices. In what follows, we will
present an alternative derivation of this result which provides 
additional insight into the localisation process and offers improved 
estimates for the localisation time.

\section{Introduction of $U_{\lambda}$ and the reduced space 
${\cal{H}}^{\prime}$}
\label{reduced space}

We first note that the operator $U'$ is close to $U$ in the sense that we can 
write $U^{\prime}=U-2U\left|sv\right\rangle\left\langle sv\right|$. The 
additional term $2U\left|sv\right\rangle\left\langle sv\right|$ changes only 
a few entries in $U$. In fact, we can choose a basis where $U$ and $U^{\prime}$ 
are identical in all but one entry.

We may thus regard the additional term as a localised perturbation of $U$.
In order to study how this perturbation effects the spectrum, we consider a
family of operators $U_{\lambda}$ changing continuously from 
$U$ to $U^{\prime}$ as $\lambda$ is varied from 0 to 1. The following
definition
\begin{equation}
\label{Ulam}
U_{\lambda}=U\left(\mathbbm{1}_{2^{n}n}+\left(\e^{\i\lambda\pi}-
1\right)\left|sv\right\rangle \left\langle sv\right|\right)
\end{equation}
fulfils this condition, that is, $U_\lambda$ is a continuous matrix valued
function in $\lambda$ and equals $U$ and $U^{\prime}$ for $\lambda=0$ and 
$\lambda=1$, respectively. $U_{\lambda}$ is in addition periodic  
with period 2 and is unitary for all $\lambda$.

In order to understand the effect of the perturbation on the spectrum 
of $U'$, we consider the symmetries of the hypercube; they 
can be described in terms of two types
of symmetry operations, $P_{i}$ and $P_{ij}$ with $i,j=1,\ldots n$.  
Writing the set of vertices as $n$-digit strings containing $0$s and $1$s, 
$P_{i}$ is defined as the operation that flips the $i$th digit 
from $0 \to 1$ or $1\to0$, respectively,
and $P_{ij}$ is the operator exchanging the $i$th and the $j$th
digit. Note that $P_{i}$ changes the Hamming weight of the vertex whereas
$P_{ij}$ does not.  The group of symmetry operations on the hypercube 
is generated by $P_{i}$ and  $P_{ij}$. Both operators represent
reflections at an $n-1$ dimensional manifold  orthogonal to 
$\vec{e}_{i}$ and $\vec{e}_{i}-\vec{e}_{j}$ respectively, where 
$\vec{e}_{i}$ is defined as the unit vector pointing in the 
$i$th direction.

Let us assume that the marked vertex sits at $v=\vec{o}$. Such a 
perturbation breaks all symmetries created by $P_{i}$. The marked 
vertex (with coin $C_1$) is, however, at a fixed point of $P_{ij}$ 
and the corresponding symmetries are not affected by the perturbation. 
As a result, not all symmetries of the unperturbed spectrum are 
lifted and the analysis can be done in a symmetry reduced space.

In particular, all eigenvectors of $U$ orthogonal to 
$\left|sv\right\rangle$ are also eigenvectors of $U_{\lambda}$ and 
their corresponding eigenvalues remain 
unchanged when varying $\lambda$ in (\ref{Ulam}). Since these 
eigenvectors are not affected by the perturbation introduced by the marked 
vertex, we concentrate our investigation on eigenvectors that are not 
orthogonal to $\left|sv\right\rangle$. We reorganise the eigenvectors 
such that there is only one eigenvector in each degenerate eigenspace 
which is not orthogonal to $\left|sv\right\rangle$.  These vectors 
are given by
$\left|\omega^{\pm\prime}_{k}\right\rangle 
=\sum_{\vec{l}, \left|\vec{l}\right|=k} 
\left|v^{\pm}_{\vec l}\right\rangle\left\langle 
v^{\pm}_{\vec l}\mid sv\right\rangle$
which after normalisation yields
\begin{eqnarray} \label{red-basis}
\left|\omega^{\pm}_{k}\right\rangle
:=\frac{1}{ \sqrt{n\choose k}}\sum_{\vec{l} 
\atop \left|\vec{l}\right|=k} \left(-1\right)^{\vec{l}\cdot\vec{v}}
\left|v^{\pm}_{\vec l}\right\rangle
,\end{eqnarray}
where $\vec{v}$ are the coordinates of the marked vertex $v$; this 
definition is not independent of $v$. In the following, we will assume 
that the marked vertex is at $\vec 0$. We discuss the general case 
in Sec.\ \ref{sec:orgspac}. We then obtain
\begin{equation} \label{eigenfred}
\left|\omega^{\pm}_{k}\right\rangle=\frac{1}{ \sqrt{n\choose k}}
\sum_{\vec{l} \atop \left|\vec{l}\right|=k} 
\left|v^{\pm}_{\vec l}\right\rangle
\end{equation}
as the set of $2n$ normalised eigenvectors of $U$ containing the marked vertex 
$v$. These vectors span a $2n$-dimensional subspace ${\cal{H}}^{\prime}$ of the 
full Hilbert space \cite{SKW03}. Note, that the marked state 
$\left| sv\right\rangle$ is in ${\cal{H}}^{\prime}$ and that the trivial 
eigenvectors are orthogonal to $\left|sv\right\rangle$ \cite{ref11}. 
Thus, ${\cal{H}}^{\prime}$ is mapped onto itself under 
the map $U_\lambda$.  That implies, that the definition of $U_\lambda$ in 
Eqn.\ (\ref{Ured}) holds for the reduced space, 
\begin{equation}
\label{Ured}
U_{\lambda}=U\left(\mathbbm{1}_{2{n}}+
\left(\e^{\i\lambda\pi}-1\right)
\left|sv\right\rangle \left\langle sv\right|\right)
 \end{equation}
where $U$, $U_{\lambda}$ are the quantum walks restricted to the reduced 
space in the basis (\ref{eigenfred}).

\begin{figure}
\centering
\includegraphics[scale=0.50]{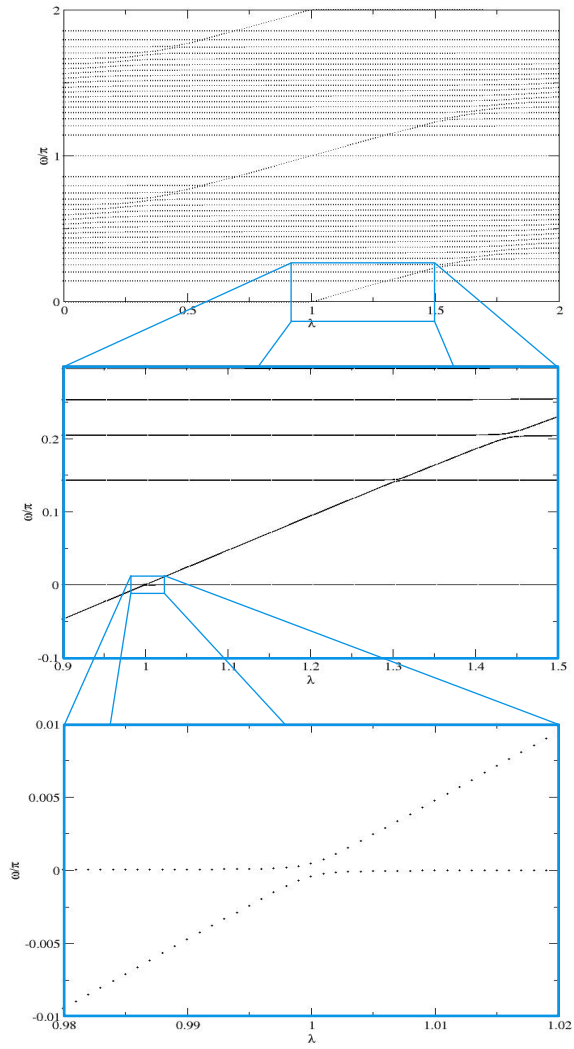}
\caption{Phases of the eigenvalues of $U_{\lambda}$ in units of $\pi$ as 
a function of $\lambda$ for a $20$ dimensional hypercube in the 
reduced space.}
\label{fig:eigenphases}
\end{figure}

\section{Spectrum of $U_{\lambda}$}
\label{sec:spec}

We start by describing main features of the spectrum of $U_\lambda$.  
Fig.~\ref{fig:eigenphases} shows 
the eigenphases of the unitary matrix $U_{\lambda}$ as a function 
of $\lambda$ in the $2 n$ dimensional reduced space ${\cal{H}}^{\prime}$. 
To simplify notation, we define a new index $m$ replacing $k$ and $\pm$ such 
that $m\in\{-n+1,n\}$ and $\{k,\pm\}=\{\left|m\right|,
{\rm sgn}\left(m\right)\}$. We
furthermore write the eigenvalues and eigenvectors as 
$\e^{\pm\i\omega_{k}} =\e^{\i\omega_{m}}$ and 
$\left|\omega_{k}^{\pm}\right\rangle =\left|\omega_{m}\right\rangle$,
respectively.

The numerical results indicate that the eigenphases $\omega_m$
of the unperturbed walk, corresponding to $\lambda = 0$ or 2 and 
given in Eqn.\ (\ref{eigenw}), remain largely unchanged when changing 
$\lambda$. In addition, there are ``perturber'' states 
with eigenphases roughly parallel to the line $\frac{\pi}{2} \lambda$. 
One finds avoided crossings at points where the eigenphases related
to $\omega_m$ `cross' the perturber states.  In 
the following we will concentrate on the perturber state 
$|u_\lambda \rangle$ causing an avoided crossing at $\omega=0$ and 
$\lambda=1$; this is called the  $m =0$-th crossing.

The dynamics at an avoided crossing can essentially be described in terms
of a two level system where the interaction induced by the map $U_\lambda$ 
between the states at the crossing is much larger than that with any of 
the other unperturbed eigenstates. At the $m$-th crossing at 
$\lambda = \lambda_{m}$, say, we can construct a two-level dynamics 
between the unperturbed eigenvector $|\omega_m\rangle$ given in 
(\ref{eigenfred}) and a perturber state $ |u_{m}\rangle$ to be 
determined below.  These two states have 
a large overlap with the exact eigenvectors $|w_m^\pm \rangle$ at 
$\lambda = \lambda_m$.  We can in fact write the eigenvectors in good 
approximation as 
$|w_m^\pm\rangle \approx (|\omega_m\rangle \pm |u_{m}\rangle)/\sqrt{2}$
for a suitable choice of phases.

Performing the walk in the two dimensional subspace spanned by 
$\left|\omega_{m} \right\rangle$ and $\left| u_{m}\right\rangle$ 
makes it possible to rotate the start state $\left|\omega_{m}\right\rangle$
into the target state $\left| u_{m}\right\rangle$.  The latter has a 
large overlap with the state of the marked vertex $\left |sv\right\rangle$ 
and its nearest neighbours.
(It will be shown that 
$\left\langle sv\mid u_{m}\right\rangle\approx 2^ {-1/2}$).

The time it takes to perform the rotation from $\left|\omega_{m}
\right\rangle$ to $\left| u_{m}\right\rangle$ by applying 
$U_\lambda$ at $\lambda = \lambda_{m}$ is determined by the gap $\Delta_m$ 
between the two levels at the $m$-th avoided crossing. One finds
\begin{eqnarray}
U^{t}_{\lambda_{m}}\left|\omega_{m}\right\rangle
&=&\frac{1}{\sqrt{2}}\left(\e^{\i t\left(\omega_{m}+
\frac{\Delta_{m}}{2}\right)}\left|w_{m}^+
\right\rangle + \e^{\i t\left(\omega_{m}-
\frac{\Delta_{m}}{2}\right)}\left|w_{m}^-\right\rangle 
\right)
\, .
\end{eqnarray}
When choosing the time $t = T_m$, such that
$\e^{\i T_m\frac{\Delta_{m}}{2}}=\i$, that is,
\begin{equation} \label{time}
T_m=\frac{\pi}{\Delta_{m}} \, ,
\end{equation}
we have
\begin{equation}
U^{T_m}_{\lambda_{m}}\left|\omega_{m}\right\rangle
=\e^{\i T_m\omega_{m}}\i \left|u_{m}\right\rangle \, .
\end{equation}
In practise, $T_m$ is the nearest integer to $\frac{\pi}{\Delta_{m}}$.
This procedure is very much in analogy to Grover's algorithm 
\cite{Gro96, NC00} except that the relation between the exact eigenstates
and the start and target states is only an approximation here. Note, that
increasing the gap $\Delta_m$ leads to a speed up of the search. 

\begin{figure}
\centering
\includegraphics[scale=0.40]{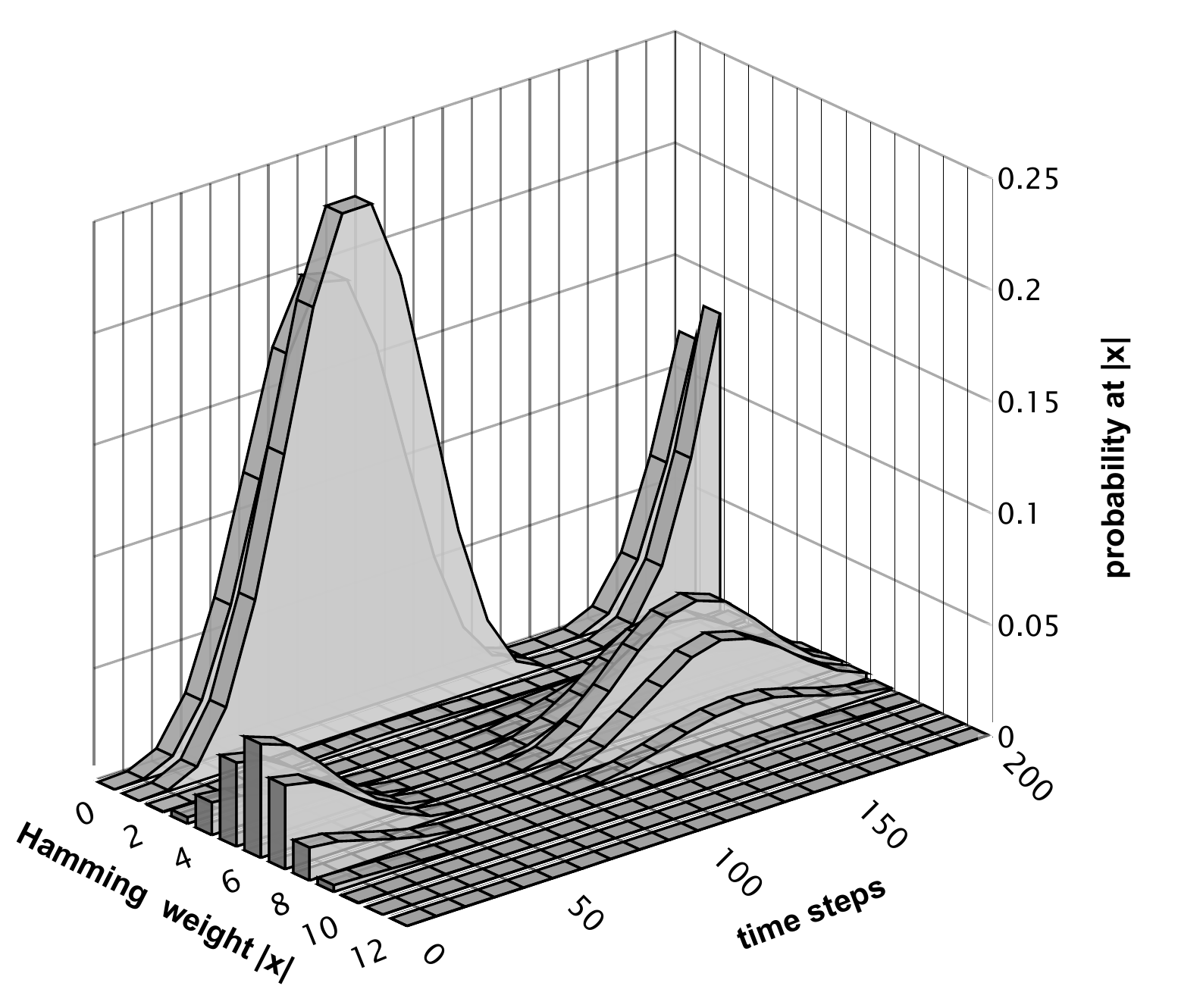}
\caption{Performance of the search algorithm. The search algorithm on the 
$n=12$ dimensional hypercube in the reduced space; (all vertices with the 
same Hamming weight merge into one point). At $t=0$ the walk starts in 
the state $|\omega_0\rangle$ corresponding to the uniform distribution
in the full space and localises at the marked vertex $v=\vec{0}$ at 
$t=74$.} \label{fig:simulation}
\end{figure}

The quantum search algorithm \cite{SKW03} works
at $\lambda = 1$, that is, at the $m=0$ crossing starting with 
the initial distribution $\left|\omega_{0}\right\rangle$; it localises at the 
marked vertex after ${\cal O}(\sqrt{N})$ time steps where $N=2^{n}$ is the 
number of vertices of the hypercube. Note that $\left|\omega_{0}\right\rangle$ 
corresponds to the uniform distribution in the full space.
The quantum walk for $n=12$ and $m=0$ is shown in Fig.\ \ref{fig:simulation}
starting on $\left|\omega_{0}\right\rangle$. One clearly sees a strong
localisation at the marked vertex $|0\rangle$ after roughly $70$ steps.

The origin of the gap
is discussed at length in \cite{SKW03} ; we see here that it emerges
through an avoided crossing. In fact, every avoided crossing can 
potentially be exploited as a search algorithm in the reduced space. 

\section{Approximate eigenvectors and eigenvalues of $U_{\lambda}$}

We will now derive an approximation for the perturber
state $|u_\lambda\rangle$ as well as the spectral gaps $\Delta_m$.  
Starting from the definition of $U_\lambda$ and $|\omega_m\rangle$ in 
Eqs.\ (\ref{eigenfred}), (\ref{Ured}),  one obtains for 
\begin{equation} \label{unpert}
U_{\lambda}\left| \omega_{m}\right\rangle=
\e^{i\omega_{m}}\left| \omega_{m}\right\rangle+
\left(\e^{\i\lambda\pi}-1\right)\beta_{m}\sqrt{n \choose |m|}2^{-n/2-1/2}
\e^{\varphi_m}U\left| sv\right\rangle
,\end{equation}
where $\e^{\pm\i\varphi_{k}} =\frac{\sqrt{k} \pm \i\sqrt{n-k}}{\sqrt{n}} \equiv
\e^{\i\varphi_{m}}$ and $\beta_m$ is given in (\ref{beta}). Thus, 
$\left| \omega_{m}\right\rangle$ is already an approximate eigenvector of 
$U_{\lambda}$ with exponentially small remainder term of the order 
${\cal O}(2^{-n/2})$ for $m$ close to $0$ or $n$. The eigenvalues of 
these vectors correspond to the horizontal lines in Fig.\
\ref{fig:eigenphases}

We now come to the construction of the perturber state $|u_{\lambda}\rangle$;
let $g\left(\lambda\right)$ be the eigenphase of the perturber state, 
that is, the eigenvalues of $U_\lambda$ are 
$\pm\e^{\i g\left(\lambda\right)}$. We expect that $g(\lambda)$ is roughly
given by 
$g\left(\lambda\right)\approx\left(\lambda-1\right)\frac{\pi}{2}$.
We set
\begin{eqnarray}
\label{def:v}
\left|u_{\lambda}\right\rangle
= 2^{-n/2-1}\sum_{m=-n+1}^{n} \sqrt{n\choose |m|}
\e^{\i\varphi_{m}} \beta_{m} a_{m} \left|\omega_{m}\right\rangle
\end{eqnarray}
for some yet unknown set of coefficients $a_{m}$. Writing 
$\left|sv\right\rangle$ in the $\left|\omega_{m}\right\rangle$-basis, 
one finds
\begin{eqnarray}
\left|sv\right\rangle 
=2^{-n/2-1/2}\sum_{m=-n+1}^{n} 
\sqrt{n\choose |m|}\e^{\i\varphi_{m}} 
\beta_{m} \left|\omega_{m}\right\rangle
.\end{eqnarray}
The scalar product $\left\langle sv\mid u_{\lambda}\right\rangle$ gives
\begin{eqnarray} \label{matrix}
\left\langle sv\mid u_{\lambda}\right\rangle=2^{-n-3/2}b
,\end{eqnarray}
where $b$ is defined as 
\begin{equation} \label{defb}
b=\sum_{m=-n+1}^{n} {n\choose |m|}a_{m}\beta_{m}^{2} 
.\end{equation}

The aim is to construct $\left| u_{\lambda}\right\rangle$ such that 
$U_{\lambda}\left|u_{\lambda}\right\rangle =
\e^{\i g\left(\lambda\right)}\left|u_{\lambda}\right\rangle$. 
Using the representation (\ref{def:v}), we obtain
\begin{eqnarray}
U_{\lambda}\left|u_{\lambda}\right\rangle
&=&U\left|u_{\lambda}\right\rangle+\left(\e^{\i\lambda\pi}-1\right)
U\left|sv\right\rangle \left\langle sv\mid u_{\lambda}\right\rangle
			\nonumber \\
&=&\e^{\i g\left(\lambda\right)}\left|u_{\lambda}\right\rangle 
+2^{-n/2-1}\sum_{m=-n+1}^{n} \sqrt{n\choose |m|}
\e^{\i\varphi_{m}}\beta_{m}\left|\omega_{m}\right\rangle \nonumber \\
\label{expansion}
& &\left[\left(-\e^{\i g\left(\lambda\right)} +e^{\i\omega_{m}}\right) a_{m}
+\left(\e^{\i\lambda\pi}-1\right)2^{-n-1}e^{\i\omega_{m}}b\right] 
.\end{eqnarray}
We can thus choose
\begin{equation}
\label{a_m}
a_{m}=\frac{\left(1-\e^{\i\lambda\pi}\right)2^{-n-1}b\, 
\e^{\i\omega_{m}}}{\e^{\i\omega_{m}}-\e^{\i g\left(\lambda\right)}} \, ,
\end{equation}
so that the second part of Eqn.\ (\ref{expansion}) vanishes.
Note that $b$ still depends on the coefficients $a_{m}$ and Eqn.\
(\ref{a_m}) represents thus a homogeneous set of coupled linear 
equations. A solution of this set of equations exists only if 
the determinant of the coefficient matrix vanishes. 
Inserting (\ref{a_m}) in (\ref{defb}) we may write this 
condition in terms of a `sum rule'
\begin{equation}
\label{summenformel}
\frac{2^{n+1}}{1-\e^{\i\lambda\pi}}=\sum_{m=-n+1}^{n}{n \choose |m|} 
\frac{\beta_{m}^{2}\e^{\i\omega_{m}}}
{\e^{\i\omega_{m}}-\e^{\i g\left(\lambda\right)}}
,\end{equation}
which implicitly defines the eigenphases $g(\lambda)$. The coefficient
$b$ thus remains undetermined. So far, 
we have only rewritten the eigenvalue equation and we thus 
expect $n$ different solutions for $g(\lambda)$ for every value 
of $\lambda$.  Note, that the singular behaviour whenever
$\omega_m \approx g(\lambda)$ indicates that the corresponding
coefficient $a_m$ dominates the expansion (except near an 
avoided crossing or for the perturber states).

For the corresponding eigenvector, we find
\begin{eqnarray}
 \left|u_{\lambda}\right\rangle
&=& 2^{-n/2-1}\sum_{m=-n+1}^{n} \sqrt{n\choose |m|}
\e^{\i\varphi_{m}} a_{m} \left|\omega_{m}\right\rangle \nonumber \\
&=&b\,2^{-\frac{3}{2} n-2}\left(1-\e^{\i\lambda\pi}\right)
\sum_{m=-n+1}^{n} \sqrt{n\choose |m|}
\frac{\e^{\i\varphi_{m}+\i\omega_{m}}\beta_{m}}
{\e^{\i\omega_{m}}-\e^{\i g\left(\lambda\right)}} \left|\omega_{m}\right\rangle
,\end{eqnarray}
and we identify $b$ as a normalisation constant.
Note, that the vector $\left|u_{\lambda}\right\rangle$ coincides with the 
basis vectors $\left|\omega_{m}\right\rangle$ for $\lambda=2j$, 
$j\in\mathbbm{Z}$ and $g(\lambda)\approx \omega_m$. 

We are here mostly interested in finding the states forming the two level 
system at an avoided crossing. The corresponding eigenspace at the $m$-th crossing
will be spanned by the unperturbed eigenstate $\left|\omega_{m}\right\rangle $
(which is already exponentially close to the true eigenvalue, see (\ref{unpert}))
and a second approximate eigenvector. This second vector may be found  
by defining a local vector near the $m$-th crossing $\left|u_{m}\right\rangle$ orthogonal
to $\left|\omega_{m}\right\rangle $; we set 

\begin{equation}
\label{u_k}
 \left|u_{m}\right\rangle
= b\, 2^{-\frac{3}{2} n-2}\left(1-\e^{\i\lambda\pi}\right)
\sum_{l=-n+1\atop l\neq m}^{n} \sqrt{n\choose |l|}
\frac{\e^{\i\varphi_{l}+\i\omega_{l}}\beta_{l}}
{\e^{\i\omega_{l}}-\e^{\i g\left(\lambda\right)}} \left|\omega_{l}\right\rangle 
.\end{equation}
If $b\ll 2^{3/2 n}$, we can assume that the neglected term is small and 
$\left|u_{m}\right\rangle$ is a good local approximation to the
true eigenvalue. An estimate for $b$ verifying this assumption
will be given below.

The compatibility condition (\ref{summenformel}) then takes on the form
of a local sum rule (setting $a_l =0$), 
\begin{equation}
\label{lokale summenformel}
\frac{2^{n+1}}{1-\e^{\i\lambda\pi}}=\sum_{l=-n+1\atop l\neq m}^{n}{n \choose |l|} 
\frac{\e^{\i\omega_{l}}\beta_{l}^2}{\e^{\i\omega_{l}}-\e^{\i g\left(\lambda\right)}}
.\end{equation}

We can use (\ref{lokale summenformel}) to obtain local approximations 
of the phase $g(\lambda)$ 
near the $m$-th crossing. We will concentrate here on the main 
crossing at $\lambda=1$ for $m$=0. Neglecting interaction with the unperturbed
state $\left|\omega_{0}\right\rangle$ (taken into account in the next section), 
we set $\e^{\i g\left(1\right)}=1$ at the crossing, that is, we demand 
$g\left(1\right)=0$.

We will show that
\begin{equation} \label{sum}
S(\lambda) := \frac{1 - \e^{\i\pi\lambda}}{2^{n+1}} 
\sum_{l=-n+1\atop l\neq 0}^{n} {n \choose |l|} 
\frac{\e^{\i\omega_l}\beta_{l}^{2}}{\e^{\i\omega_l}-
\e^{\i g\left(\lambda\right)}}=1 + {\cal O}(e^{-n})
\end{equation}
at $\lambda = 1$, that is, Eqn.\ (\ref{lokale summenformel}) is 
fulfilled up to an exponentially small error term. 

Using that the spectrum is symmetric with respect to $0$, that is,
for every $l\in[-n+1,n]$ there exists one $k\in[-n+1,n]$ such that 
$\e^{\i\omega_k}=-\e^{\i\omega_l}$ and 
writing $S(\lambda)$ in terms of $\omega_l \in (0,\pi)$ only, we obtain
\begin{equation}
S(\lambda) = \frac{1 - \e^{\i\pi\lambda}}{2^{n}} 
\sum_{l=1}^{n-1} {n \choose l} 
\frac{\e^{2 \i\omega_l}}{\e^{2 \i\omega_l}-\e^{\i 2g\left(\lambda\right)}} 
+\frac{1 - \e^{\i\pi\lambda}}{2^{n+1}} 
.\end{equation}
Using $\sum_{l=0}^n{n \choose l}  = 2^n$, we may write
\begin{eqnarray}
\label{def:g_lambda}
S(\lambda) &=& \frac{1 - \e^{\i\pi\lambda}}{2^{n+1}} \left[ 2^n -1
- \i\sum_{l=1}^{n-1} {n \choose l} 
\cot\left(\omega_l -g\left(\lambda\right)\right)\right]
.\end{eqnarray}

Setting $g\left(1\right)=0$, we find $S(1)= 1+2^{-n}$. 
By expanding $S(\lambda)$ in a power series in $\lambda$ and demanding 
that the derivatives of $S(\lambda)$ vanish at $\lambda=1$, we obtain 
conditions for the derivatives of $g\left(\lambda\right)$; in particular, 
one finds, 

\begin{eqnarray}
g^{\prime}\left(1\right)=\frac{\pi}{2\left(1+\gamma_n-2^{-n+1}\right)}
\end{eqnarray}

with
\begin{eqnarray}
\gamma_n =  \frac{1}{2^{n}}
\sum_{l=1}^{n-1} {n \choose l} 
\cot^2(\omega_l) \sim \frac{1}{n} \quad \mbox{for large } n
.\end{eqnarray}
The last estimate is obtained asymptotically by using the 
de Moivre-Laplace theorem and Poisson summation.
Note, that for 
$n\rightarrow \infty$, this result coincides with 
$g\left(\lambda\right)\approx (\lambda-1) \pi/2$ as stated in the 
beginning of the section. Similarly, 
we can construct functions $g(\lambda)$ at crossings $m\ne 0$.

From Eqn.\ (\ref{lokale summenformel}), we determine
the normalisation constant $b$ by writing
\begin{eqnarray}
1&=&\left\langle u_{m}\mid u_{m}\right\rangle \nonumber \\
&=& \left|b\right|^{2} 2^{-3n-4}\left|1-\e^{\i\pi\lambda}\right|^{2} 
\sum_{l=-n+1\atop l\neq m}^{n} {n\choose |l|}
\frac{\beta_{l}^{2}}{\left|
\e^{\i\omega_{l}}-\e^{\i g\left(\lambda\right)}\right|^{2}}
\end{eqnarray}
and thus
\begin{eqnarray}
\frac{1}{ \left|b\right|^{2} }
&=&2^{-3n-4}\left|1-\e^{\i\pi\lambda}\right|^{2} 
\sum_{l=-n+1\atop l\neq m}^{n} {n\choose |l|}
\frac{\beta_{l}^{2}}{\left|\e^{\i\omega_{l}}-
\e^{\i g\left(\lambda\right)}\right|^{2}} \, .
\end{eqnarray}
The sum can be calculated using Eqn.\ (\ref{lokale summenformel}) 
(and neglecting exponentially small terms). The derivative of 
(\ref{lokale summenformel}) with respect to $\lambda$ leads to
\begin{equation}
\frac{\pi}{g^{\prime}\left(\lambda\right)}\frac{2^{n+1}}
{\left|1-\e^{\i\lambda\pi}\right|^{2}}
=\sum_{l=-n+1\atop l\neq m}^{n}{n \choose |l|}
\frac{\beta_{l}^{2}}{\left|\e^{\i\omega_{l}}-
\e^{\i g\left(\lambda\right)}\right|^{2}}
,\end{equation}
and hence $\left|b\right|^{2} =
\frac{2g^{\prime}\left(\lambda\right)}{\pi}2^{2n+2}$. 
We are free to choose a phase for the vector $|u_m\rangle$ and set
\begin{equation} \label{b-est}
b :=\sqrt{\frac{2g^{\prime}\left(\lambda\right)}{\pi}}2^{n+1}
.\end{equation}
Therefore the condition $b \ll 2^{3n/2}$ stated after Eqn.\ (\ref{u_k}) is 
fulfilled for large $n$ and $\left|u_{m}\right\rangle$ is an approximative 
eigenvector of $U_{\lambda}$. 

\section{Matrix elements of $U_{\lambda}$.}

To calculate the size of the gap $\Delta_m$ 
between the two eigenvalues at the $m$-th avoided crossing, 
we consider how  $U_\lambda$ acts on the $2\times 2$ space 
spanned by $|\omega_m\rangle$ and 
$|u_m\rangle$.  At the $m$th crossing, we find 
$\lambda=\lambda_{m}=g^{-1}\left(\omega_{m}\right)$  
and the corresponding matrix elements of $U_\lambda$
are 
\begin{eqnarray*} 
A&:=&\left\langle u_{m}\mid U_{\lambda_{m}} \mid u_{m} \right\rangle
= \e^{\i g\left(\lambda_{m}\right)}		\\
B&:=&\left\langle \omega_{m}\mid U_{\lambda_{m}}\mid  
u_{m} \right\rangle\\
&=& 2^{-n/2-1}\left(\e^{\i\lambda_{m}\pi}-1\right)  
\e^{\i\omega_{m}+\i\varphi_{l}} \sqrt{{n \choose |m|}} 
\beta_{m}\sqrt{\frac{2g^{\prime}\left(\lambda\right)}{\pi}}
\\
C&:=&\left\langle u_{m}\mid U_{\lambda_{m}}\mid  
\omega_{m} \right\rangle\\
&=& -2^{-n/2-1}\left(\e^{-\i\lambda_{m}\pi}-1\right)  
\e^{\i g\left(\lambda_{m}\right)-\i\varphi_{m}}\sqrt{{n \choose |m|}} 
\beta_{m}\sqrt{\frac{2g^{\prime}\left(\lambda\right)}{\pi}}
	\\
D&:=&\left\langle \omega_{m}\mid U_{\lambda_{m}}
\mid \omega_{m}  \right\rangle
=\e^{\i\omega_{m}}\left(1+ 
\left(\e^{\i\lambda_{m}\pi}-1\right)2^{-n-1} 
\beta_{m}^{2}{n \choose |m|}\right)	
.\end{eqnarray*}
The  $2\times 2$ matrix $W = \begin{pmatrix} A & B \\ C & D \end{pmatrix}$ 
characterises the avoided crossings in $U_{\lambda}$ to a good
approximation. 
Note that we have 
$|A| \sim |D| \sim 1$ and $|B| \sim |C| \sim 
\sqrt{{n \choose |m|}} e^{-n/2-1} \ll 1$
for small $m/n$.  

The gap $\Delta_{m}$ in the spectrum is 
given by the difference of the two eigenphase; for small 
differences, this is in leading order given by the difference between 
the two eigenvalues of $W$.

We thus obtain
\begin{eqnarray}
\Delta_{m}^{2
}&:=& \left(D-A\right)^{2} + 4BC \nonumber \\
&=&\left(\e^{\i\omega_{m}}-\e^{\i g\left(\lambda_{m}\right)}+2^{-n-1}
{n \choose |m|} \beta_{m}^{2} \left(\e^{\i\lambda_{m}\pi}-1\right)
\e^{\i\omega_{m}}\right)^{2}   \nonumber \\
& &- 2^{-n}\left|1-\e^{\i\lambda_{m}\pi}\right|^{2} 
{n \choose |m|} \e^{\i g\left(\lambda_{m}\right)+\i\omega_{m}} 
\beta_{m}^{2}\, \frac{2g^{\prime}\left(\lambda\right)}{\pi} \, .
\end{eqnarray}

By construction, we have $g\left(\lambda_{m}\right)=\omega_{m}$
at the crossing, and thus
\begin{equation} \label{luecke-f}
\left|\Delta_{m}\right|=\left|1-\e^{\i\lambda_{m}\pi}\right| 
2^{-n/2}\sqrt{n \choose |m|} \beta_{m}\sqrt{\frac{2g^{\prime}
\left(\lambda\right)}{\pi}} 
+{\cal O}\left(2^{-n}{n \choose |m|}\right)
,\end{equation} 
neglecting terms of the order $2^{-n}$ both in the matrix elements 
and in the 'sum rule' (\ref{sum}).
\begin{figure}
   \centering
\includegraphics[scale=0.30]{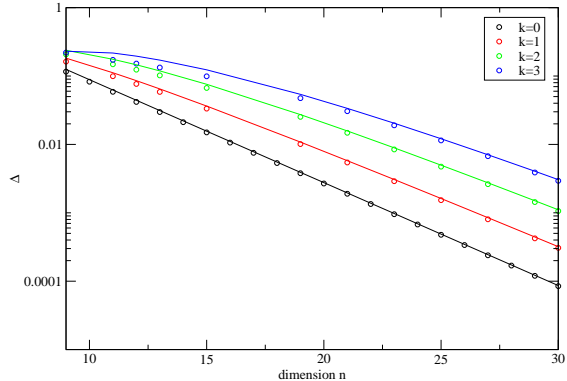}
\caption{Comparison of numerical and theoretical values for the size 
of the gap. The circles correspond to numerical results while the 
solid lines correspond to the theoretical results given by 
$2^{-n/2}\sqrt{{n \choose |m|}}\left|1-\e^{\i\lambda_{m}\pi}\right|$ 
for several values $m$ as a function of the dimension $n$. }
  \label{fig:luecke}
\end{figure}

In Fig.~\ref{fig:luecke} we compare the leading order term, Eqn.\ 
(\ref{luecke-f}), with numerical results for the first four crossings 
as a function of the dimension of the hypercube. Clearly, our 
estimate captures the behaviour of the gap very well, even at 
intermediate values of $n$ down to $n \approx 10 - 15$. We note
in particular that the size of the gap increases with the order
of the crossing $|m|$.

\section{Time of the search}
\label{Zeit}

The quantum algorithm takes place in the two dimensional subspace 
spanned by the 
two approximate eigenvectors involved in the avoided crossing, 
$\left|\omega_{m}\right\rangle$ and $\left| u_{m}\right\rangle$. 
In analogy to Grover's algorithm \cite{Gro96, NC00}, the search
corresponds to a rotation from an initial state $|\omega_m\rangle$
to a final state localised at the marked item and its nearest
neighbours; see the discussion in Sec.\ \ref{sec:spec}. 
Direct calculation using (\ref{matrix}) and (\ref{b-est}) yields, 

\begin{equation} 
\left\langle u_{m}\mid sv\right\rangle\approx2^{-1/2} 
\sqrt{2g^{\prime}\left(\lambda\right)/\pi} \approx
\left|\left\langle u_m\mid U\mid sv\right\rangle\right|
,\end{equation} 
where $\sqrt{2g^{\prime}\left(0\right)/\pi}\approx 1$ for large $n$
and we approximate $u_\lambda$ by $u_m$ at the crossing.
That is, the target state $\left|u_{m}\right\rangle$ has a large 
overlap with the marked vertex $|sv\rangle$ or its immediate 
neighbours, $U|sv\rangle$. Any measurement will thus yield with a 
high probability either $|sv\rangle$ or one of its neighbouring vertices when
done after $T_m$ time steps. This implies, 
that we can define a quantum search algorithm for {\em every} avoided
crossing $m$ (as long as the dynamics effectively takes place
in a  two-level system). The search time $T_m$ is given by 
Eqn.\ (\ref{time}), that is 
\begin{equation} \label{time-m}
T_{m}=\frac{\pi 2^{n/2}\sqrt{\frac{\pi}{2g^{\prime}\left(\lambda_m\right)}}}
{\left|1-\e^{\i\lambda_{m}\pi}\right|\sqrt{n \choose |m|} \beta_{m}}
.\end{equation} 
In particular, the $T_m$ decreases with increasing $m$ thus 
making search algorithms at higher order crossings potentially 
more effective. 

\begin{figure}
\centering
\includegraphics[scale=0.35]{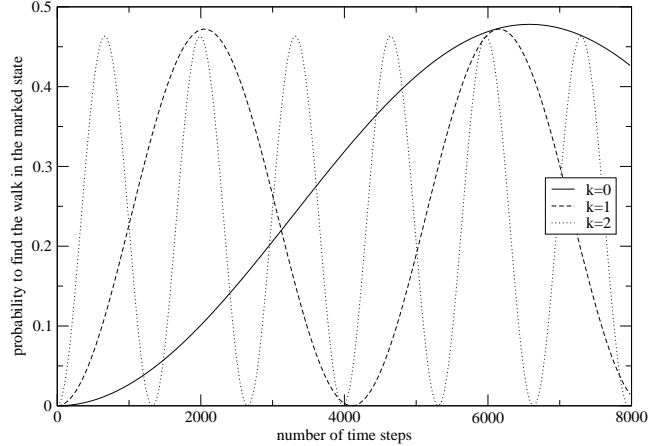}
\caption{The probability to measure the state at the marked vertex 
in $n=25$ dimensions for the central crossing $m = 0$ and the 
first ancillary crossings $m =1$ and $m = 2$.}
\label{fig:nebenkreuzungen}
\end{figure}

Fig.~\ref{fig:nebenkreuzungen} 
shows the probability to measure the quantum walk in the reduced space at 
the marked vertex as a function of time and $n = 25$. 
It is evident from the figure, that the quantum search localises on 
the marked vertex with a 50\% probability for all algorithms shown 
(corresponding here to the crossings $m =0, 1, 2$). In particular, 
the search becomes faster with increasing $m$ in accordance with
(\ref{time-m}). For the search algorithm for $m=2$, we find the marked 
vertex in $1/10$th of the time compared to a search with $m=0$ and only a 
small loss in amplitude. If we proceed to higher $m$, one looses more 
and more amplitude and the search becomes inefficient; in our example 
($n = 25$) this happens for  $m>4$.
In general, we find that the quantum search algorithm fails to localise 
at the marked vertex when the gap is of the size of the distance between 
two adjacent (unperturbed) eigenvalues.

\section{Results in the original space $\cal H$ for general $v$}
\label{sec:orgspac}

The class of search algorithms considered in the previous section
act in the reduced space finding the marked vertex in
${\cal O}\left(\sqrt{N}\right)$ time. To start the search, 
one needs to know the parameter values $\lambda_m$ and the 
initial state $|\omega_m\rangle$. The former can in principle 
be obtained to arbitrary accuracy for a given dimension $n$
or are known explicitly as in the case $m=0$.  The starting vectors 
$|\omega_m\rangle$ defined in Eqn.\ (\ref{red-basis}) depend, however,
on the choice of the marked vertex $v$ for $m \ne 0$. In fact, 
the reduction of the space itself as described in Sec.\ 
\ref{reduced space} is not independent of the marked vertex; when 
simplifying Eqn.\ (\ref{red-basis}) to Eqn.\ (\ref{eigenfred}), we 
explicitly put the marked vertex at $v=\vec{0}$. Any other choice
of $v$ will, however, change the starting vectors $|\omega_m\rangle$ 
for $m\ne 0$ or equivalently change the point of origin for the symmetry
operations $P_i$, $P_{ij}$ (see Sec., \ref{reduced space}).

We are ultimately interested in search algorithms on the 
full hypercube. This can be achieved directly by
employing the $m=0$ crossing. The corresponding initial
vector $|\omega_0\rangle$ is independent of $v$ with $\lambda_0 = 1$; 
this yields the search algorithm in 
\cite{SKW03}. It is, however, the slowest algorithm 
of the ones discussed in the previous section. To
make use of any of the other search algorithms, we need
the starting vector $|\omega_m\rangle$ given in Eqn.\ 
(\ref{red-basis}), which depends, however, on the marked vertex
itself. 

This optimal starting vector is embedded in the ${ n \choose |m|}$ 
dimensional vector space related to the eigenvalue $m$ in the 
unperturbed space $\cal{H}$. This space is spanned by the vectors 
$|v_{\vec k}^{\pm}\rangle $, 
(where $\pm\vert\vec{k}\vert=m$), given in Eqn.\ (\ref{eigenv}).
So, in addition to finding the marked vertex, one also needs to search
for the state $|\omega_m\rangle$.  We have not succeeded in devising an 
efficient method for finding this optimal starting vector for arbitrary $v$,
that is, a method that would not wipe out any gains made by 
improving the search time $T_m$ in Eqn.\ (\ref{time-m}). 

We conclude that the algorithms introduced above have 
search times which are all of the same order in the reduced 
space; only the original algorithm devised in \cite{SKW03} is,
however, useful for searches on the full hypercube  as no extra efforts are 
needed in finding both the values for $\lambda_m$ andi, more importantly,
the optimal starting vector $|\omega_m\rangle$. The technique
presented here provides an improved estimate for the search time and offers a 
new point of view by studying quantum random walks in terms of avoided
crossings.  \\[.5cm]
{\bf Acknowledgements}:\\
We thank Fritz Haake and Brian Winn for carefully reading the 
manuscript and for valuable comments.\\[.5cm]


\begin{thebibliography}{1}
\bibitem{SKW03} N\,Shenvi, J\,Kempe and K\,B\,Whaley: {\it Quantum 
random-walk search algorithm} {\it Physical Review A}, {\bf 67} 052307
(2003).

\bibitem{Gro96} L Grover: {\it A fast quantum mechanical algorithm for 
database search} in {\it Proc. 28th STOC},
ACM Press, Philadelphia, Pennsylvania, p.\ 212 (1996);
L K Grover, {\it Quantum mechanics helps in searching for a 
needle in a haystack} {\it Physical Review Letters}, {\bf 97} 325 (1997).

\bibitem{NC00} M A Nielsen and I L Chuang, {\it Quantum Computation and Quantum 
Information} {\it Cambridge University Press} (2000).

\bibitem{kempe} J\,Kempe: {\it Quantum random walks - 
an introductory overview}, {\it Contemporary Physics}, {\bf 44} 307 (2003)

\bibitem{Amb03} A Ambainis, {\em Quantum walks and their algorithmic 
applications}, {\it International Journal of Quantum Information}, 
{\bf 1} 507, (2003).

\bibitem{SS06} S Gnutzmann and U Smilansky, {\it Quantum Graphs: 
Applications to quantum chaos and universal spectral statistics}, 
{\it Advances in Physics} {\bf 55}, 527 (2006).

\bibitem{ST04} S\ Severini and G\ Tanner, {\em Regular Quantum Graphs}, 
{\it J.\ Phys.\ A} {\bf 37} 6675 (2004).

\bibitem{Tan07} G\,Tanner: {\it From quantum graphs to quantum random walks}
in {\it Non-Linear Dynamics and Fundamental Interactions} Springer, 
Dordrecht, p 69 (2006).

\bibitem{AA03} S Aaronson and A Ambainis, 
{\em Quantum search of spatial regions}, {\it Proc. 44th Annual IEEE Symp.\ on 
Foundations of Computer Science (FOCS)}, p.\ 200 (2003).

\bibitem{AKR05} A\,Ambainis, J\,Kempe and A\,Rivosh: {\it 
Coins make quantum walks faster} {\it Proc. 16th ACM-SIAM SODA}, 
p.\ 1099 (2005)

\bibitem{CG04} A M Childs and J Goldstone, {\it Spatial search by quantum walk}
{\it Physical Review E}, {\bf 70} 022314 (2004).

\bibitem{ref11} C Moore and A Russell, in {\it Proceedings of RANDOM, 2002}, 
edited by J\,D\,P\,Rolim and P\,Vadham, Springer, Cambridge, MA, 
p.\ 164, (2002).


\end{thebibliography}
\end{document}